\documentclass[pra,twocolumn]{revtex4}

\usepackage{amsfonts}
\usepackage{amssymb}
\usepackage{amsbsy}
\usepackage{bm}

\begin{document}


\title{Fresnel coefficients as hyperbolic rotations}
\author{J. J. Monz\'on and L. L. S\'anchez-Soto}
\affiliation{Departamento de \'{O}ptica, 
Facultad de Ciencias F\'{\i}sicas, 
Universidad Complutense, 28040 Madrid, Spain}

\date{\today}

\begin{abstract}
We describe the action of a plane interface
between two semi-infinite media in terms of
a transfer matrix. We find a remarkably simple
factorization of this matrix, which enables us
to express the Fresnel coefficients as a 
hyperbolic rotation.

\end{abstract}

\maketitle

\narrowtext

\section{Introduction}

Reflection and transmission at a discontinuity are
perhaps the first wavelike phenomena that one 
encounters in any undergraduate physics course.
The physics underlying this behavior is well
understood: mismatched impedances generate 
the reflected and transmitted waves, while the 
application of the proper boundary conditions at 
the discontinuity  provide their corresponding amplitude 
coefficients~\cite{CR68}. Moreover, this general 
framework facilitates a unified treatment for all 
the types of waves appearing in Nature.

For light waves the impedance is proportional to 
the refractive index. Accordingly, the behavior 
of light at the plane interface  between two 
semi-infinite media are derived in most optics 
textbooks~\cite{BO99,HE99,PE87}. The 
resulting amplitude coefficients are described 
by the famous Fresnel formulas. It seems almost 
impossible to say anything new about these 
Fresnel  formulas. However, a quick look at 
the indexes of  this  Journal~\cite{AmJPhys}, 
among others~\cite{others1,others2}, 
immediately reveals a steady flow of papers 
devoted to subtle aspects of this problem, 
which shows that the topic is far richer than one 
might naively expect.

In this paper we reeleborate once again 
on this theme. We present the action of any
interface in terms of a transfer matrix,
and we find a hitherto unsuspectedly simple
factorization of this matrix. After renormalizing
the field amplitudes, such a factorization leads
us to introduce a new parameter in terms of 
which Fresnel formulas appear as a hyperbolic 
rotation.

As our teaching experience demonstrates, the 
students have troubles in memorizing Fresnel
formulas due to their fairly complicated appearance.
In this respect, there are at least two reasons that,
in our opinion, bear out the interest of our contribution:
first, Fresnel formulas appears in the new variables
as a hyperbolic rotation that introduces remarkable 
simplicity and symmetry. Second,  this formalism 
is directly linked to other fields of physics, mainly 
to special relativity, which is more than a 
curiosity~\cite{VG93,MO99e,MO01}.

\section{The interface transfer matrix}

Let two homogeneous isotropic semi-infinite 
media, described by complex refractive indices 
$N_0$ and $N_{1}$, be separated 
by a plane boundary. The $Z$ axis is chosen 
perpendicular to the boundary and directed as 
in Fig.~1.

We assume an incident monochromatic, linearly 
polarized plane wave from medium 0, which 
makes an angle $\theta _{0}$ with the $Z$ axis 
and has amplitude $E_{0}^{(+)}$. The electric field 
is either in the plane of incidence (denoted by superscript
$\parallel$) or perpendicular to the plane of incidence 
(superscript $\perp$). This wave splits into a  reflected 
wave $E_{0}^{(r)}$ in medium 0, and a transmitted 
wave $E_{0}^{(t)}$ in medium 1 that makes and 
angle $\theta _{1}$ with the $Z$ axis. 
The angles of incidence $\theta_0$ and refraction
$\theta_1$ are related by the Snell's law 
\begin{equation}
N_0 \sin \theta _0 = N_1 \sin \theta _1.  
\end{equation}
If media 0 and 1 are transparent (so that $N_0$ and 
$N_1$ are  real numbers) and no total reflection 
occurs, the angles $\theta_0$ and  
$\theta_1$ are also real and the above picture of 
how a plane wave is reflected and refracted at the 
interface is simple. However, when either one or 
both media is absorbing, the angles  $\theta_0$ and  
$\theta_1$  become, in general, complex and the 
discussion continues to hold only  formally, but
the physical picture of the fields becomes 
complicated~\cite{ST63}.

We consider as well another plane wave of the 
same frequency and polarization, and amplitude  
$E_{1}^{(-)}$, incident from medium $1$ at an angle  
$\theta _{1}$, as indicated in Fig.~1. In the same way, 
we shall denote by $E_{1}^{(r)}$ and $E_{1}^{(t)}$ 
the reflected and transmitted amplitudes of the 
corresponding waves.

The complex amplitudes of the total output fields at 
opposite points immediately above and below  the 
interface will be called  $E_{0}^{(-)}$ and 
$E_{1}^{(+)}$, respectively. The wave vectors of 
all waves lie in the plane of incidence and when the 
incident fields are $\parallel$ or $\perp$ polarized, 
all  plane waves excited by the incident ones have 
the same polarization.

The amplitudes $E_{0}^{(-)}$ and $E_{1}^{(+)}$
are then given by
\begin{eqnarray}
\label{E1}
E_{0}^{(-)} & = & E_{0}^{(r)}+E_{1}^{(t)}=
r _{01}E_{0}^{(+)}+t _{10}E_{1}^{(-)},  \nonumber \\
& & \\
E_{1}^{(+)}& = & E_{0}^{(t)}+E_{1}^{(r)}=
t _{01}E_{0}^{(+)}+r _{10}E_{1}^{(-)}, \nonumber 
\end{eqnarray}
where $r _{01}$ and $t _{01}$ are the Fresnel  reflection
and transmission coefficients for the interface 01, and  
$r_{10}$ and $t _{10}$ refer to the corresponding 
coefficients for the interface 10. These Fresnel 
coefficients are determined by demanding that across 
the boundary the tangential components of $\mathbf{E}$ 
and $\mathbf{H}$ should be continuous~\cite{BO99}.
For nonmagnetic media they are given by
\begin{eqnarray}
r _{01}^{\parallel} & = & \frac{N_1 \cos\theta_0 -N_0 \cos\theta_1}
{N_1 \cos\theta_0 +N_0 \cos\theta_1},  \nonumber \\
& & \\ \nonumber
t _{01}^{\parallel} & = &  \frac{2N_0 \cos\theta_0}
{N_1 \cos\theta_0 +N_0 \cos\theta_1},    \nonumber \\
& & \nonumber \\
& & \nonumber \\
r_{01}^\perp & = & \frac{N_0 \cos\theta_0 -N_1 \cos\theta_1}
{N_0 \cos\theta_0 +N_1 \cos\theta_1},  \nonumber \\
& & \\ 
t _{01}^\perp  & = &  \frac{2N_0 \cos\theta_0}
{N_0 \cos\theta_0 +N_1 \cos\theta_1} , \nonumber
\end{eqnarray}
for  both basic polarizations. It is worth noting that,  
although these equations are written for electromagnetic 
waves, it is possible to translate all the results for particle-wave 
scattering, since there is a one-to-one correspondence
between the propagation in an interface between two 
media of  electromagnetic waves and of the nonrelativistic
particle waves satisfying Schr\"{o}dinger equation~\cite{LE87}.

The linearity revealed by Eqs.~(\ref{E1}) suggests the use
of $2 \times 2$ matrix methods. However, Eqs.~(\ref{E1}) 
links output to input fields, while the standard way of 
treating this topic is by relating the field amplitudes at 
each  side of the interface. Such a relation is expressed  
as~\cite{AZ87,MO96}
\begin{equation}
\left( 
\begin{array}{c}
E_{0}^{(+)} \\ 
E_{0}^{(-)}
\end{array}
\right ) 
= 
\mathsf{I}_{01}
\left( 
\begin{array}{c}
E_{1}^{(+)} \\ 
E_{1}^{(-)}
\end{array}
\right) .
\end{equation}
The choice of these column vectors is motivated from
the optics of layered media, since it is the only way of
calculating the field amplitudes at each side of every
layer by an ordered product of matrices. 

We shall call $\mathsf{I}_{01}$ the interface transfer 
matrix and, from Eqs.~(\ref{E1}), is given by
\begin{equation}
\mathsf{I}_{01} = \frac{1}{t _{01}}
\left ( 
\begin{array}{cc}
1 & -r _{10} \\ 
r _{01} \ \  & t _{01}t _{10}-r _{01}r _{10}
\end{array}
\right ) .  \label{interface1}
\end{equation}

By using a matrix formulation of the boundary 
conditions~\cite{YE88} one can factorize the 
interface transfer matrix $\mathsf{I}_{01}$  
in the new and remarkable form~\cite{others2} 
(that otherwise one can also check directly using 
the Fresnel formulas)
\begin{eqnarray}
\label{matricesps}
\mathsf{I}_{01}^\parallel  & = &  
\mathsf{R}^{-1}(\pi /4) 
\left ( 
\begin{array}{cc}
\cos \theta _{1}/\cos \theta _{0} & 0 \\ 
0 & N_{1}/N_0
\end{array}
\right )
\mathsf{R}(\pi /4) ,  \nonumber  \\
& & \\
\mathsf{I}_{01}^\perp & = & 
\mathsf{R}^{-1}(\pi /4)
\left ( 
\begin{array}{cc}
(N_{1}\cos \theta _{1})/(N_0\cos \theta _{0}) \ \ & 0 \\ 
0 &  1
\end{array}
\right )
\mathsf{R}(\pi /4) ,   \nonumber 
\end{eqnarray}
where
\begin{equation}
 \mathsf{R}(\pi /4) = 
\frac{1}{\sqrt{2}}
\left ( 
\begin{array}{rr}
1 & -1 \\ 
1 & 1
\end{array}
\right ) 
\end{equation}
represents the matrix of a clockwise rotation of angle 
$\pi /4$. Now, it is straightforward to convince oneself 
that a diagonal matrix postmultiplied  by $\mathsf{R}(\pi /4)$ 
and premultiplied by its inverse is always of the form
\begin{equation}
\left ( 
\begin{array}{cc}
a & b \\ 
b  & a
\end{array}
\right )  ,
\end{equation}
$a$ and $b$ being, in general,  complex numbers. This 
result implies that $\mathsf{I}_{01}$ in Eq.~(\ref{interface1}) 
must be also of this form, which turns out the constraints 
\begin{eqnarray}
\label{Stokes}
r _{10} & = & -r _{01},  \nonumber \\
& & \\
1 + r _{01}r _{10} & = & t _{01}t _{10} . \nonumber   
\end{eqnarray}
This applies to both basic polarizations by  the simple 
attachment of a label to all the coefficients and 
constitutes an alternative algebraic demonstration 
of the well-known Stokes relations without resorting 
to the usual  time-reversal argument~\cite{LE87,YE88}. 
Similar results can be also derived  in particle scattering 
from the unitarity requirement  on the \textit{S} matrix. 
However, note that  the equality $|r_{10}|  = | r_{01} |$ 
implied by  Eq.~(\ref{Stokes}) can become 
counter-intuitive  when applied to particle reflection, 
since one might  expect stronger reflection for particle 
waves moving  up in a potential gradient than for those 
going down. In fact, these relations, as emphasized by 
Lekner~\cite{LE87}, ensure that the reflectivity 
is exactly the same in the two cases, unless there 
is total internal reflection.  

In summary, these Stokes relations allows one to
write~\cite{AZ87,MO96}
\begin{equation}
\mathsf{I}_{01}=\frac{1}{t _{01}}
\left ( 
\begin{array}{cc}
1 & r _{01} \\ 
r _{01} & 1
\end{array}
\right ) .  
\label{interface2}
\end{equation}

It is worth noting that the inverse matrix satisfies
$\mathsf{I}_{01}^{-1} = \mathsf{I}_{10}$ and 
then describes the interface taken in the reverse
order. The physical meaning of these matrix 
manipulations is analyzed in Section~\ref{meaning}.

\section{Renormalization of field amplitudes}

From Eqs.~(\ref{matricesps}) one directly obtain that, 
for both basic polarizations, we have
\begin{equation}
\det  \mathsf{ I}_{01}^\parallel = 
\det \mathsf{I}_{01}^\perp = 
\frac{N_1 \cos \theta_1}{N_0 \cos \theta_0} \neq 1.
\end{equation}
For the reasons that will become clear in  
Section~\ref{hyperbolic},  it is  adequate to 
renormalize the field amplitudes  to ensure 
that the transfer matrix has always unit  
determinant. To this end, let us define
\begin{eqnarray}
\label{campnor}
e^{(\pm)}_0 = \sqrt{N_0 \cos \theta_0}\ 
E^{(\pm)}_0, \nonumber \\
& & \\
e^{(\pm)}_1 = \sqrt{N_1 \cos \theta_1} \
E^{(\pm)}_1 .  \nonumber 
\end{eqnarray}
Accordingly, the action of the interface is 
described now by
\begin{equation}
\label{normact}
\left( 
\begin{array}{c}
e_{0}^{(+)} \\ 
e_{0}^{(-)}
\end{array}
\right ) 
= 
\mathsf{ i}_{01}
\left( 
\begin{array}{c}
e_{1}^{(+)} \\ 
e_{1}^{(-)}
\end{array}
\right) ,
\end{equation}
where the renormalized interface matrix is
\begin{eqnarray}
\label{interface3}
\mathsf{i}_{01}&=&\mathsf{R}^{-1}(\pi /4)
\left ( 
\begin{array}{cc}
1/\xi _{01}& 0 \\ 
0 & \xi_{01}
\end{array}
\right )
\mathsf{R}(\pi /4) \nonumber \\
& & \nonumber \\
& = & \frac{1}{2}
\left ( 
\begin{array}{cc}
\xi_{01} + 1/\xi_{01}\ \ &  \xi_{01} - 1/\xi_{01}  \\ 
\xi_{01} - 1/\xi_{01} \ \  & \xi_{01} + 1/\xi_{01}
\end{array}
\right )  ,
\end{eqnarray}
and the factor $\xi_{01}$ has  the values
\begin{eqnarray}
\xi_{01}^\parallel  & = & \sqrt{\frac{N_1 \cos \theta_0}
{N_0 \cos \theta_1}} = 
\sqrt{\frac{\sin(2\theta_0)}{\sin(2\theta_1)}}, \nonumber \\
& &  \\
\xi_{01}^\perp  & =  & \sqrt{\frac{N_0 \cos \theta_0}
{N_1 \cos \theta_1}} =
\sqrt{\frac{\tan \theta_1}{\tan \theta_0}}. \nonumber
\end{eqnarray}
Other way of expressing these relations is
\begin{eqnarray}
\xi_{01}^\parallel \xi_{01}^\perp & =  & 
\frac{\cos \theta_0}{\cos \theta_1}, \nonumber \\
&&  \\
\frac{\xi_{01}^\parallel}{\xi_{01}^\perp} &  =  & 
\frac{N_1}{N_0}. \nonumber 
\end{eqnarray}

It is now evident from Eq.~(\ref{interface3})
that the renormalized interface matrix 
satisfies $\det \mathsf{i}_{01} = +1$, as desired.
Moreover, by taking into account the general form
given in Eq.~(\ref{interface2}), we can reinterpret 
$\mathsf{i}_{01}$ in terms of renormalized
Fresnel coefficients as
\begin{equation}
\label{interfasehat}
\mathsf{i}_{01} =
\frac{1}{\widehat{t}_{01}}
\left ( 
\begin{array}{cc}
1 & \widehat{r}_{01} \\ 
\widehat{r}_{01} & 1
\end{array}
\right ) , 
\end{equation}
where
\begin{eqnarray}
\label{reltxi}
\widehat{r}_{01}  &=& 
\frac{\xi_{01}- 1/\xi_{01}}
{\xi_{01}+ 1/\xi_{01}} , \nonumber \\
& &  \\
\widehat{t}_{01} &=& \frac{2 }
{\xi_{01}+ 1/\xi_{01}}, \nonumber
\end{eqnarray}
which satisfy
\begin{equation}
\widehat{r}_{01}{}^2 + \widehat{t}_{01}{}^2 =  1 .
\end{equation}
This relation does not trivially reduce to the 
conservation of the energy flux on the interface, 
because the complex reflection and transmission 
coefficients appear in the form $\widehat{r}_{01}{}^2$ 
and $\widehat{t}_{01}{}^2$ instead of  
$|\widehat{r}_{01}|^2$ and $| \widehat{t}_{01}|^2$.
In fact, it can be seen as a consequence of the 
renormalization factors appearing in the definition
(\ref{campnor}) that project the direction of the
corresponding wave vector onto the normal to the
boundary.

The Fresnel coefficients can be obtained from the
renormalized ones as
\begin{eqnarray}
r_{01}  &=&\widehat{r}_{01}, \nonumber \\ 
& &  \\
t_{01}  & = & \sqrt{\frac{N_1 \cos \theta_1} 
{N_0 \cos \theta_0}} \ \widehat{t}_{01}. \nonumber
\end{eqnarray}

It is clear from Eqs.~(\ref{reltxi}) that the single 
parameter $\xi_{01}$ gives all the information about 
the interface, even for absorbing media or when total
reflection occurs.  We have $\mathsf{i}_{01}^{-1}  = 
\mathsf{i}_{10}$; that is, the inverse also describes 
the interface taken in  the reverse order. Thus,  
$\xi_{10} = 1/\xi_{01}$ and it follows that 
\begin{eqnarray}
\widehat{r}_{01} &=& - \widehat{r}_{10}, \nonumber \\
& & \\
\widehat{t}_{01} &=& \widehat{t}_{10}.  \nonumber 
\end{eqnarray}

In Fig.~2 we have plotted  the behavior of 
$\xi_{01}^\parallel$ and  $\xi_{01}^\perp$ 
as a function of the angle of incidence $\theta_0$, for an
interface air-glass ($N_0/N_1 = 2/3$) and, for the
purpose of comparison, the corresponding values of 
$r_{01}^\parallel $ and $r_{01}^\perp$. The discussion
about these amplitude coefficients and the corresponding
phase shifts can be developed much in the same way
as it is done in most of the undergraduate optics 
textbooks.

\section{The interface as a hyperbolic rotation}
\label{hyperbolic}

The definition of the renormalized transfer matrix 
for an interface in Eq.~(\ref{interface3}) may appear,
at first sight, rather artificial. In this Section
we shall interpret its meaning by recasting it in 
an appropriate form that will reveal the origin of 
the rotation matrices $\mathsf{R}(\pi/4)$.

To simplify as much as possible the
discussion, let us assume that we are
dealing with an interface between two
transparent media when no total 
reflection occurs. In this relevant case,
the Fresnel reflection and transmission
coefficients, and therefore $\xi_{01}$, are
real numbers. Let us introduce a new
parameter $\zeta_{01}$ by 
\begin{equation}
\xi_{01} = \exp(\zeta_{01}/2) .
\end{equation}
Then, the action of the interface can be
expressed as
\begin{equation}
\label{girohiper}
\left (  \begin{array}{c}
e_{0}^{(+)} \\ 
e_{0}^{(-)}
\end{array}
\right)
 =
\left ( 
\begin{array}{cc}
\cosh (\zeta_{01}/2)  & \sinh (\zeta_{01}/2) \\ 
\sinh (\zeta_{01}/2) &  \cosh (\zeta_{01}/2)
\end{array}
\right )
\left (  
\begin{array}{c}
e_{1}^{(+)} \\ 
e_{1}^{(-)}
\end{array}
\right ) ,
\end{equation}
where the renormalized Fresnel  coefficients 
can be written now  as
\begin{eqnarray}
\widehat{r}_{01} &=& \tanh (\zeta_{01}/2), \nonumber \\ 
& & \\
 \widehat{t}_{01} &=& 
\frac{1}{\cosh (\zeta_{01}/2)}. \nonumber 
\end{eqnarray}

Given the importance of this new reformulation 
of the action of an interface, some comments seem 
pertinent: it is clear that the reflection coefficient 
can be always expressed as a hyperbolic tangent,
whose addition law is simple.  In fact, such an 
important result was first derived by Khashan~\cite{KH79} 
and  is the origin of  several approaches for treating the 
reflection coefficient of layered structures~\cite{CO91}, 
including bilinear or quotient functions~\cite{LI98}, 
that  are just of the form (\ref{reltxi}).  However, 
the transmission coefficient for these structures seems 
to be (almost)  safely ignored in the literature, because 
it behaves as a hyperbolic secant, whose addition law 
is more involved.

Now the meaning of the rotation $\mathsf{R}(\pi/4)$
can be put forward in a clear way.
To this end, note that the transformation 
(\ref{girohiper}) is formally a hyperbolic 
rotation of angle $\zeta_{01}/2$ acting on the complex 
field variables $[e^{(+)}, e^{(-)}]$. As it is usual in
hyperbolic geometry ~\cite{CO69}, it is convenient
to study this transformation in a coordinate 
frame whose new axes are the bisecting 
lines of the original one. In other words, 
in this frame whose axes are rotated
$\pi/4$ respect to the original one,
the new coordinates are
\begin{equation}
\left(  
\begin{array}{c}
\tilde{e}^{(+)} \\
\tilde{e}^{(-)}
\end{array}
\right ) =
\mathsf{R}(\pi/4) 
\left(  
\begin{array}{c}
e^{(+)} \\
e^{(-)}
\end{array}
\right )  
\end{equation}
for both 0 and 1 media, and the action 
of the interface is represented by the matrix
\begin{equation}
\left( 
\begin{array}{c}
\tilde{e}_{0}^{(+)} \\ 
\tilde{e}_{0}^{(-)}
\end{array}
\right ) 
= 
\left ( 
\begin{array}{cc}
1/\xi _{01}& 0 \\ 
0 & \xi_{01}
\end{array}
\right )
\left ( 
\begin{array}{c}
\tilde{e}_{1}^{(+)} \\ 
\tilde{e}_{1}^{(-)}
\end{array}
\right ) ,
\end{equation}
which is a squeezing matrix that scales 
$\tilde{e}_1^{(+)}$ down to the factor 
$\xi _{01}$ and  $\tilde{e}_1^{(-)}$
up by the same factor.

Furthermore, the product of these complex 
coordinates remains constant 
\begin{equation}
\tilde{e}_{0}^{(+)} \tilde{e}_{0}^{(-)} =
\tilde{e}_{1}^{(+)} \tilde{e}_{1}^{(-)} ,
\end{equation}
or  
\begin{equation}
\label{invariante2}
[e_{0}^{(+)} ]^2  -  [e_{0}^{(-)}  ]^2  =
[e_{1}^{(+)} ]^2  -  [e_{1}^{(-)}  ]^2 ,
\end{equation}
which appears as a fundamental invariant of any
interface. In these renormalized field variables
it is nothing but the hyperbolic invariant of the 
transformation. When viewed in the original field 
amplitudes it reads as 
\begin{equation}
N_0 \cos \theta _{0} \{ [ E_{0}^{(+)}]^2 - 
[E_{0}^{(-)} ] ^2 \} = 
N_1 \cos \theta _{1} \{ [ E_{1}^{(+)}]^2 -
[ E_{1}^{(-)} ] ^2 \} ,
\label{invariante}
\end{equation}
which was assumed as a basic axiom 
by Vigoureux and Grossel~\cite{VG93}. 

To summarize this discussion at a glance, 
in  Fig.~3 we have plotted the unit hyperbola
$[e^{(+)} ]^2  -  [e^{(-)}  ]^2 = 1$, assuming
real values for all the variables. The interface
action transforms then the point 1 into the 
point 0. The same hyperbola, when referred
to its proper axes, appears as  $\tilde{e}^{(+)} 
\tilde{e}^{(-)} = 1/2$.

\section{The physical meaning of interface 
composition}
\label{meaning}

To conclude, it seems adequate to provide a
physical picture of the matrix manipulations we
have performed in this paper. First, the inverse 
of an interface matrix, as pointed out before, 
describes the interface taken  in the reverse order. 

Concerning the product of interface matrices,
this operation has physical meaning only when
the second medium of the first interface is identical 
to the first medium of the second one. In this case, 
let us consider the interfaces 01 and 12. A direct
calculation from Eqs.~(\ref{matricesps}) shows
that
\begin{equation}
\mathsf{I}_{01}\mathsf{I}_{12}=\mathsf{I}_{02},
\end{equation}
for both basic polarizations, which is equivalent 
to the constraints 
\begin{eqnarray}
\label{Lorentz}
r _{02} &=&\frac{r _{01}+ r _{12}}
{1+ r _{01}r _{12}}, \nonumber \\
&& \\
t_{02}&=&\frac{t _{01}t _{12}}
{1+ r _{01} r _{12}}. \nonumber 
\end{eqnarray}
Note that the reflected-amplitude composition 
behaves as a $\tanh$ addition law, just as in the famous
Einstein addition law for collinear velocities:
no matter what values the reflection amplitudes
$r_{01}$ and $r_{12}$ (subject only to 
$| r_{01}| \le 1$ and $| r_{12} | \le 1$) have, 
the modulus of the composite amplitude  $| r_{02} |$)
cannot exceed the unity. Alternatively, we have
\begin{equation}
\widehat{r}_{02} = r_{02} = \tanh (\zeta_{02}/2) = 
 \tanh (\zeta_{01}/2 + \zeta_{12}/2) 
\end{equation}
which leads directly to the first one of  Eqs.(\ref{Lorentz}).

On the contrary, the transmitted amplitudes composes
as a  sech, whose addition law is more involved and is
of little interest for our purposes here. 

Obviously, for this interface composition to be realistic
one cannot neglect the wave propagation between 
interfaces. However, this is not an insuperable
drawback. Indeed, let us consider a  single layer of a
transparent material of refractive index $N_1$ and 
thickness $d_1$ sandwiched between two semi-infinite 
media 0 and 2. Let
\begin{equation}
\beta _{1}=\frac{2\pi }{\lambda }N_1 d_1
\cos \theta _1 
\end{equation}
denote the phase shift due to the propagation in the 
layer, $\lambda$ being  the wavelength in vacuum. 
A standard calculation gives for the reflected and 
transmitted amplitudes by this layer the Airy-like
functions~\cite{BO99} 
\begin{eqnarray}
R_{012} &=&\frac{r _{01}+r _{12}
\exp (-i2\beta_1)}{1+r _{01}r_{12}\exp (-i2\beta_1)},  
\nonumber \\
&& \\
T_{012} &=&\frac{t _{01}t _{12}\exp (-i\beta_1)}
{1+r _{01}r_{12}\exp (-i2\beta _{1})}.  \nonumber
\end{eqnarray}
The essential point is that in the limit $\beta_1 = 2 n \pi $
($n = 0, 1, \ldots $),  which can be reached either 
when $d_1 \rightarrow 0$ or when the plate is under
resonance conditions, then $R_{012}\rightarrow
r _{02}$\ and $T_{012}\rightarrow t_{02}),$ and we 
recover Eqs.~(\ref{Lorentz}). This gives perfect sense
to the matrix operations in this work.

\section{Conclusions}

We have discussed in this paper a simple transformation
that introduces remarkable simplicity and symmetry in
the physics of a plane interface. In these new suitable 
variables the action of any interface appears in a natural way 
as a hyperbolic rotation, which is the natural arena
of special relativity. 

This formalism does not add any new physical ingredient
to the problem at hand, but allows one to obtain previous
results (like Fresnel formulas or Stokes relations) in 
a particularly simple and elegant way that appears
closely related to other fields of physics.

\begin{figure}
\caption{Wave vectors of the incident, reflected, 
and transmitted fields at the interface 01.}
\end{figure}

\begin{figure}
\caption{Plot of the factor $\xi_{01}$ and $r_{01}$ 
as functions of the angle of incidence $\theta_0$ 
(in degrees) for both  basic polarizations for 
an interface air-glass ($N_0 = 1,  N_1= 1.5$). 
The marked points correspond to the Brewster
angle.}
\end{figure}

\begin{figure}
\caption{Schematic plot of the hyperbolic rotation
performed by the interface 01 that transforms on
the unit hyperbola the point 1 into the point 0.}
\end{figure}

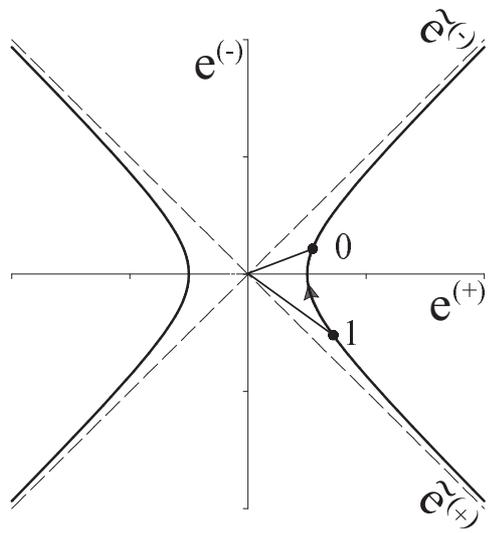
\end{document}